\documentclass[prb,nopacs,twocolumn,preprintnumbers,amsmath,amssymb]{revtex4}
\usepackage[latin1]{inputenc}
\usepackage{epsfig}
\usepackage{graphicx}
\usepackage{dcolumn}
\usepackage{color}
\usepackage{bm}

\begin{document}

\title{Elliot-Yafet mechanism in graphene}

\author{H. Ochoa$^1$, A. H. Castro Neto$^{2,3}$, F. Guinea$^1$ }

{\affiliation{$^1$ Instituto de Ciencia de Materiales de Madrid. CSIC. Sor Juana In\'es de la Cruz 3. 28049 Madrid. Spain. \\
$^2$ Graphene Research Centre and Physics Department, National University of Singapore, 2 Science Drive 3, 117542, Singapore. \\
$^3$ Department of Physics, Boston University, 590 Commonwealth Ave., Boston MA 02215, USA.}

\begin{abstract}
The differences between spin relaxation in graphene and in other materials are discussed. For relaxation by scattering processes, the Elliot-Yafet mechanism,
the relation between the spin and the momentum scattering times acquires a dependence on the carrier density, which is independent of the scattering mechanism and the relation between mobility and carrier concentration. This dependence puts severe restrictions on the origin of the spin relaxation in graphene. The density dependence of the spin relaxation allows us to distinguish between  ordinary impurities and defects which modify locally the spin-orbit interaction.
\end{abstract}

\maketitle
{\em Introduction.}
Graphene is considered as a potential material for spintronics devices due to the weak spin-orbit (SO) interaction\cite{HuertasHernandoetalprb} and long spin lifetimes\cite{Tombrosetal}. One of the most intriguing features of spintronics in graphene is that the observed spin relaxation time is significantly shorter than the values estimated theoretically. A number of recent experiments\cite{HanKawakami,Yangetal,Avsaretal} investigate which spin relaxation mechanism plays the major role both in single layer and in bilayer graphene. The D'yakonov-Perel\cite{DyakonovPerel} and the Elliot-Yafet\cite{Elliot,Yafet} mechanisms have been discussed in the context of graphene\cite{HuertasHernandoetalprl,Ertleretal}. Experiments suggest that the main source of spin relaxation in single layer graphene is extrinsic, lending support to the Elliot-Yafet mechanism. Longer spin lifetimes have been reported in bilayer graphene than in single layer\cite{HanKawakami}, when the SO coupling in the bilayer is expected to be somewhat stronger\cite{bilayerPaco}.

The Elliot-Yafet mechanism takes into account the change in the spin polarization of a Bloch electron due to scattering by impurities, lattice defects or phonons. The Elliot relation establishes a linear relation between the spin relaxation time and momentum scattering time:\begin{align}
\tau_s=\frac{\tau_p}{\alpha} \label{ab}
\end{align}where $\alpha$ can be interpreted as the spin-flip probability during a momentum relaxation event. Elliot deduced this relation by using a perturbative approach. Due to the SO coupling, Bloch states with well-defined spin polarization are not longer eigenstates of the complete Hamiltonian. In the case of conventional metals with a center of symmetry, two degenerate states can be defined for each value of the momentum\cite{Elliot}:
\begin{eqnarray}
\left[a_{\mathbf{k}}\left(\mathbf{r}\right)|\uparrow\rangle+b_{\mathbf{k}}\left(\mathbf{r}\right)|\downarrow\rangle\right]e^{i\mathbf{k}\cdot\mathbf{r}}\\
\left[\left(a_{-\mathbf{k}}\left(\mathbf{r}\right)\right)^*|\downarrow\rangle-\left(b_{-\mathbf{k}}\left(\mathbf{r}\right)\right)^*|\uparrow\rangle\right]e^{i\mathbf{k}\cdot\mathbf{r}}
\end{eqnarray}
where the coefficients $a$, $b$ are lattice-periodic due to the discrete translation symmetry. These two states are connected by spatial inversion and time reversal symmetries and form a Kramers' doublet. Typically these states can be identified with spin-up and spin-down states because typically $|b|\ll1$. Since the SO interaction couples electronic states with opposite spin projections in different bands (in the case of graphene the SO interaction couples $\pi$ and $\sigma$ bands), perturbation theory gives $|b|\approx\Delta_{SO} / \Delta E$, where $\Delta E$ is the energy difference between the two bands involved. Usually, $\Delta_{SO}\ll\Delta E$, as in the case of graphene.

The spin flip amplitude during the scattering by an obstacle with no spin degrees of freedom itself can be computed using the Born approximation, leading to $\alpha\approx\left\langle|b|^2\right\rangle$
where the symbol $\langle\rangle$ expresses an average over the Fermi surface. These arguments are quite general and do not depend on the nature of the scatterers. Realistic calculations can be done in some cases, for instance in the case of III-V semiconductors\cite{Chazalviel,SongKim}.

The relation (\ref{ab}) holds experimentally for most conventional metals\cite{Jedemaetalprb}. As we discuss below, doped graphene is not an exception. However, unlike ordinary metals, the nature of the effective SO coupling acting on the graphene $\pi$ electrons, which are the relevant ones in what concerns to transport properties, makes the ratio $\tau_s/\tau_p$ to depend strongly on the number of carriers through the Fermi energy. This result holds for different kind of defects, as it discussed next. A wide variety of experiments\cite{JozsaetalprbR} suggest a linear scaling between $\tau_p$ and $\tau_s$, with independence of the carrier concentration. Our analysis shows that this behavior cannot attributed to the Elliot-Yafet mechanism, opening the door to other extrinsically induced spin relaxation mechanisms, such as a defects which modify locally the spin-orbit interaction\cite{CastroNetoPaco,Weeksetal}.

{\em The model.}
In graphene, the intrinsic SO coupling can be neglected in comparison to the Rashba-like coupling, generated by perturbations which break spatial inversion, such as electric fields and ripples. If the perturbation changes slowly over scales larger than the lattice spacing, we can neglect intervalley hybridization\cite{HentschelPaco}. Then, the Hamiltonian of the problem reads:\begin{align}
\mathcal{H}=-i\hbar v_F\vec{\sigma}\cdot\nabla+\frac{\Delta}{2}\left(\vec{\sigma}\times\vec{s}\right)_z
\label{hamiltonian}\end{align}

The Rashba-like term breaks the spatial inversion symmetry, and two degenerate eigenstates cannot be defined for a given momentum $\mathbf{k}$. The Rashba-like term entangles spin and valley degrees of freedom, complicating the definition of the amount of spin relaxation in a scattering event. The Bloch eigenstates of \eqref{hamiltonian} read:\begin{align}
\Psi_{\mathbf{k},\pm}=\left[\left(\begin{array}{c}
1 \\
\frac{\epsilon_{\mathbf{k} \pm}}{\hbar v_F | \mathbf{k} |} e^{i\theta_{\mathbf{k}}}\end{array}\right)\otimes|\uparrow\rangle \pm \right. \nonumber \\ \left. \pm i  \left(\begin{array}{c} \frac{\epsilon_{\mathbf{k} \pm}}{\hbar v_F | \mathbf{k} |} e^{i\theta_{\mathbf{k}}}\\
e^{2i\theta_{\mathbf{k}}}\label{eigenstates}\end{array}\right)\otimes|\downarrow\rangle\right]
e^{i\mathbf{k}\cdot\mathbf{r}}
\end{align}
where $\theta_{\mathbf{k}}=\arctan\left(k_y/k_x\right)$ and
$\epsilon_{\mathbf{k},\pm}^e = \pm\frac{\Delta}{2} + \sqrt{(\hbar v_F | \mathbf{k} |)^2+\left(\frac{\Delta}{2}\right)^2}$, where $e$ denotes electrons. A similar expression can be defined for holes by changing the sign of the second term. In what follows, we restrict the discussion to electrons. As we see, a spin direction cannot be uniquely defined for all momenta. When we take $\Delta=0$, eigenstates \eqref{eigenstates} are Bloch states with well-defined projection of spin over the direction of motion, that is, helicity $\pm$. This is not strictly true when $\Delta\neq0$, but in the spirit of the above Elliot's approach, we can identify each of these eigenstates with chiral states $\pm$. This is justified from the point of view of perturbation theory, since for carrier concentrations of interest we have $\Delta/\epsilon_F\ll1$. Thus, the effect of the Rashba-like coupling can be interpreted as the energy splitting ($\sim\Delta$) of bands with opposite chirality.

Let's consider now scattering by a potential $U\left(\mathbf{r}\right)$ diagonal in sublattice and spin degrees of freedom in the Born approximation. We study scattering in the chiral channels discussed above instead of the spin-up and spin-down channels as in the case of the Elliot's approach. This restriction complicates the definition of the amount of spin relaxation. To illustrate this, it is useful to calculate the scattering amplitudes in these channels in the absence of SO. Assuming and incoming Bloch state with energy $\epsilon=\hbar v_Fk$ and positive chirality, it is easy to see that in that case (see Supplementary Information):\begin{eqnarray}
f_+\left(\theta\right)=-\left(\hbar v_F\right)^{-1}\sqrt{\frac{k}{8\pi}}U_{\mathbf{q}}e^{-i\theta}\left(1+\cos\theta\right)
\nonumber\\
f_-\left(\theta\right)=-\left(\hbar v_F\right)^{-1}\sqrt{\frac{k}{8\pi}}U_{\mathbf{\mathbf{q}}}ie^{-i\theta}\sin\theta
\label{amplitudes}\end{eqnarray}where $U_{\mathbf{q}}$ is the Fourier transformation of the scattering potential evaluated at the transferred momentum $\mathbf{k}'-\mathbf{k}$, and $\theta$ is the angle between the outcoming $\mathbf{k}'$ and incoming $\mathbf{k}$ momentum (see Fig.~\ref{born}). If we repeat the calculation in the spin-up and spin-down channels assuming an incoming state with spin up, then we obtain $f_{\uparrow}=-\left(\hbar v_F\right)^{-1}\sqrt{\frac{k}{8\pi}}U_{\mathbf{q}}\left(1+e^{-i\theta}\right)$ and $f_{\downarrow}=0$, since in the absence of SO there is no spin-flip. The scattering amplitude $f_-$ is not zero in general (except for forward scattering)  so that it cannot be related with a cross-section for a spin-flip process.
\begin{figure}
\includegraphics[width=0.5\textwidth]{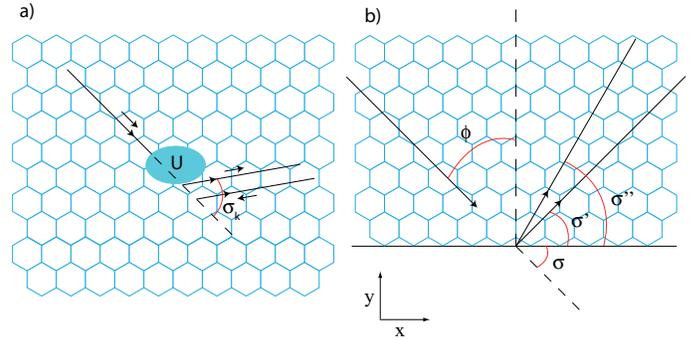}
\caption[fig]{a) Sketch of scattering by a potential $U\left(\mathbf{r}\right)$ in the chiral channels defined in the text. b) Sketch of scattering by a boundary.}
\label{born}
\end{figure}

In order to study scattering in the chiral channels defined by the Rashba coupling we define the probability for a spin-flip process from the changes in the scattering in both chiral channels due to the presence of the SO coupling. We follow the approach of Ref.~\onlinecite{HuertasHernandoetalprl} adapted to the calculation within the Born approximation. We define the quantity:\begin{align}
S\left(\theta\right)=\frac{\sum_{\pm1}\left|f_\pm^{0}\left(\theta\right)\right|\cdot\left|f_\pm^{\Delta}\left(\theta\right)-
f_\pm^{0}\left(\theta\right)\right|}{\sum_{\pm1}\left|f_\pm^{0}\left(\theta\right)\right|^2}
\label{s1}
\end{align}where the superscript $\Delta$ (0) indicates the presence (absence) of the Rashba-like coupling. This quantity vanishes when $\Delta=0$, and it can be interpreted as a measure of the amount of spin relaxed in the direction defined by $\theta$. As in the case of the Elliot's approach, the total amount of spin relaxation during a scattering event can be defined as the average of this quantity over the Fermi surface:\begin{align}
S=\left\langle S\left(\theta\right)\right\rangle=\frac{1}{2\pi}\int d\theta S\left(\theta,\epsilon=\epsilon_F\right)
\label{s2}
\end{align}

In Ref.~\onlinecite{HuertasHernandoetalprl} the relation $S\sim\Delta/\epsilon_F$ was deduced in the case of weak scatterers. Here we show that this relation is general, and it does not depend on the nature of the scatterer, including strong scatterers or other impurity potentials which cannot be treated in the Born approximation, where the value of $S$ cannot obtained from perturbation theory. The Born approximation suffices, however, to show how this behavior is implied by the nature of the SO coupling in graphene with independence of the precise scattering mechanism. It is not difficult to compute exactly $f_{\pm}^{\Delta}\left(\theta\right)$ (see Supplementary Information), but the picture provided by perturbation theory is enough to illustrate this behavior in the doped regime. For an arbitrary scatterer, the value amplitude $f_\pm^0 ( \theta )$, as defined in~\eqref{amplitudes}, requires the use of non perturbative methods. The difference $f_\pm^\Delta ( \theta ) - f_\pm ^0 ( \theta )$, however, can be obtained by expanding in powers of $\Delta / \epsilon_F$. This is easy to see in the Born approximation, where the substitution $\epsilon\rightarrow\epsilon\pm\Delta/2$ in expressions~\eqref{amplitudes} has to be made in order to obtain $f^\Delta_\pm ( \theta )$. An expansion in powers of $\Delta/\epsilon$ is well defined, and it implies that $S\left(\theta\right)\sim\Delta/\epsilon$, independently of the scattering potential, $U_\mathbf{q}$, which factorizes in expressions~\eqref{amplitudes}. Assuming this behavior, the Elliot relation for graphene can be easily found. After $N_{col}$ collisions, the change of spin polarization is of the order of $\sqrt{N_{col}}S$. Dephasing takes place after a time $\tau_s=N_{col}\tau_p$, when $\sqrt{N_{col}}S\sim 1$. Hence we obtain the relation:\begin{equation}
\tau_s\approx\frac{\epsilon_F^2}{\Delta^2}\tau_p
\label{elliot}
\end{equation}This is the Elliot relation for graphene. As one can see, the ratio $\tau_s/\tau_p$ depends on the carrier concentration through the Fermi energy. In what follows we compute exactly the amount of spin relaxation $S$ for different kind of scatterers, generalizing the relation~\eqref{elliot}.

{\em Results for different scatterers.}
The scattering amplitudes~\eqref{amplitudes} can be calculated exactly in the presence of the Rashba-like coupling (see Supplementary Information). In the case of weak scatterers, we consider as scattering center a isotropic potential $U\left(\mathbf{r}\right)=V\vartheta\left(r-R\right)$, where $\vartheta\left(r-R\right)$ is a step function. In the case of Coulomb scatterers the scattering potential reads $U\left(\mathbf{r}\right)=-\hbar v_F\alpha/r$. Note that $|\alpha|<1/2$, in other case the solutions of the Coulomb problem oscillate very fast and have no well-defined limit as $r\rightarrow0$, which corresponds to the Dirac vacuum breakdown (the continuum description in terms of the Dirac Hamiltonian is not valid) \cite{Novikov}. The results are shown in Fig.~\ref{resultados1}. In both cases $S\sim\Delta/\epsilon_F$.

\begin{center}
\begin{figure}
\includegraphics[width=0.4\textwidth]{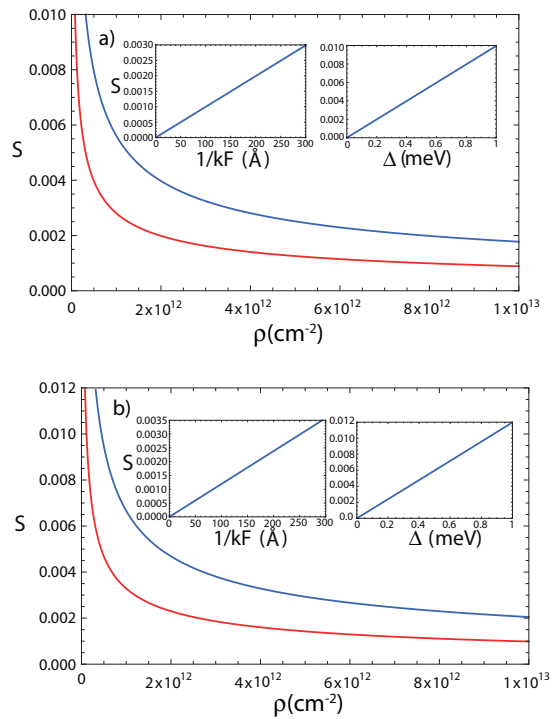}
\caption[fig]{$S$ as a function of the carrier concentration for $\Delta=1$ meV (in blue) and $\Delta=0.5$ meV (in red). a) Weak scatterers ($R=1$ {\AA} and $V_0=0.1$ eV). b) Coulomb scatterers. These results are obtained by computing exactly the scattering amplitudes $f_{\pm}^{\Delta}$ and evaluating numerically Eq.~\eqref{s2}. Insets: $S$ for $\Delta=0.1$ meV and $k_F=\left(\hbar v_F\right)^{-1}\epsilon_F=0.01$ {\AA} plotted as functions of $k_F^{-1}$ and $\Delta$ respectively. A clear linear dependence is showed, as it is argued in the text.}
\label{resultados1}
\end{figure}
\end{center}

In order to study spin relaxation during scattering by a boundary we have to adapt the definition of $S$. We are going to consider a zig-zag termination for simplicity, since it defines the most general boundary conditions\cite{AkhmerovBeenakker}. We consider as incoming wave a Bloch state $\Psi_{\mathbf{k},+}$ with energy $\epsilon$, forming an angle $\phi$ with the direction perpendicular to the boundary. As it is deduced from Fig.~\ref{born}, $\pi/2+\sigma=\phi$, where $\sigma=\arctan\left(k_y/k_x\right)$.
Two outgoing Bloch states exist satisfying conservation of energy and momentum in the direction parallel to the boundary. Then, the outgoing wave can be written as the superposition
$\Psi_{out}=r_1\Psi_{\mathbf{k}_+,+}+r_2\Psi_{\mathbf{k}_-,-}$, where $\mathbf{k}_{+}$ ($\mathbf{k}_{-}$) forms an angle $\sigma'$ ($\sigma''$) with the direction defined by the boundary (see Fig.~\ref{born}), and $\left|\mathbf{k}_{\pm}\right|\equiv k_{\pm}=\left(\hbar v_F\right)^{-1}\sqrt{\epsilon^2\mp\epsilon\Delta}$. We can define the amount of spin relaxed in the direction defined by $\phi$ as:\begin{align}
S\left(\phi\right)=\frac{\left|r_1^0\right|\cdot\left|r_1-r_1^0\right|+\left|r_2^0\right|\cdot
\left|r_2-r_2^0\right|}{\left|r_1^0\right|^2+\left|r_2^0\right|^2}
\label{sphi}
\end{align}
where the superscript $0$ refers to the reflection coefficients in the absence of the SO coupling. As before, the amount of spin relaxed by the boundary can be defined as the average, $S=\left\langle S\left(\phi\right)\right\rangle=\frac{1}{\pi}
\int_{-\pi/2}^{\pi/2}d\phi S\left(\phi,\epsilon=\epsilon_F\right)$.
By imposing zig-zag boundary conditions we obtain the following expressions for the reflection coefficients:\begin{eqnarray}
r_1=-\frac{k_-e^{i\sigma}+k_+e^{i\sigma''}}{k_-e^{i\sigma'}+k_+e^{i\sigma''}}\\
r_2=\frac{k_-\left(e^{i\sigma}-e^{i\sigma'}\right)}{k_-e^{i\sigma'}+k_+e^{i\sigma''}}
\end{eqnarray}Besides this, from conservation constrains we have $\sigma'=-\sigma$ and $\cos\sigma''=\frac{k_+}{k_-}\cos\sigma$. To first order in the SO coupling, $\sigma''=-\sigma-\frac{\Delta}{\epsilon}\cot\sigma+O\left(\Delta^2/\epsilon^2\right)$. Then, to first order in the SO coupling, the reflection coefficients in terms of the angle $\phi$ read:\begin{eqnarray}
r_1=ie^{i\phi}\sin\phi+\frac{\Delta}{2\epsilon}+O\left(\frac{\Delta^2}{\epsilon^2}\right)\\
r_2=-e^{i\phi}\cos\phi-\frac{\Delta}{2\epsilon}+O\left(\frac{\Delta^2}{\epsilon^2}\right)
\end{eqnarray}
The amount of spin relaxation can be estimated as $S=\frac{2\Delta}{\pi\epsilon_F}$. 
This expression fits the exact result rather well, see Fig.~\ref{resultados2}.

\begin{figure}
\includegraphics[width=0.5\textwidth]{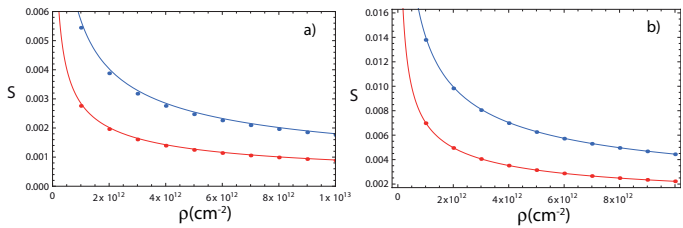}
\caption[fig]{$S$ as a function of the carrier concentration for $\Delta=1$ meV (in blue) and $\Delta=0.5$ meV (in red). The dots correspond to the numerical evaluation of $S$, the continuum line to the analytical estimates of the text. a) Scattering by boundaries. b) Strong scatterers ($R=1.4$ {\AA}).}
\label{resultados2}
\end{figure}

In the case of strong scatterers, such as vacancies, the Born approximation fails. As in the case of boundaries, we need to extend the definition of $S$. Strong scatterers can be described as a circular void of radius of the order of the lattice constant. We can exploit the cylindrical symmetry of the problem by using the decomposition of the eigenstates of \eqref{hamiltonian} into partial waves with well-defined generalized total angular momentum $J=l_z+\sigma_z/2+s_z/2$, which is actually a global symmetry of the problem, where $l_z$ is the third component of the orbital angular momentum operator $l_z=-i\left(x\partial_y-y\partial_x\right)$. For each in-coming cylindrical wave with energy $\epsilon$, there are two reflected waves with the same energy. A quantity analogous to Eq.~\eqref{sphi} can be defined, considering scattering in all channels with well-defined $J$ (see Supplementary Information). The amplitude for spin scattering can be calculated analytically $S\approx\frac{\pi\Delta}{2\epsilon_F}$
which fits very well the numerical evaluation of $S$, see Fig.~\ref{resultados2}.

The case of clusters of impurities\cite{KGG09} is studied within this formalism as well, since the Born approximation fails when the range of the scattering potential $R$ is too large in such a way that the associated energy scale $\hbar v_FR^{-1}$ exceeds its strength, $VR\ll\hbar v_F$. The same behavior $S\sim\Delta/\epsilon_F$ is deduced (see Supplementary Information).

{\em Discussion and conclusions.}
As we have seen, the averaged amount of spin relaxed during a scattering event behaves as $S\sim\Delta/\epsilon_F$, independently of the nature of the scatterer, implying the general relation $\tau_s\approx \epsilon_F^2\tau_p/\Delta^2$.
This result is not consistent with a linear scaling between the spin relaxation time and the diffusion coefficient at different gate voltages as it is observed in the experiments\cite{JozsaetalprbR,HanKawakami}, suggesting that other mechanisms dominate spin scattering. This is consistent with the fact that CVD (Chemical Vapor Deposition) graphene-based spin valves show essentially the same spin transport properties as exfoliated graphene\cite{Avsaretal}. This result implies that differences between exfoliated and CVD graphene, such as grain boundaries, do not limit spin transport. It is interesting to consider in detail the experimental data of Ref.~\onlinecite{JozsaetalprbR}. The results show a sub-linear dependence of the diffusion constant on carrier density (proportional to the momentum scattering time) and also of the spin relaxation time. This is clearly inconsistent with our result if one assumes the Elliot-Yafet induced by defects in graphene as the main spin relaxation mechanism.

The main exception to the Elliot relation comes from impurities which enhance locally the SO coupling, such as heavy impurities\cite{Weeksetal}, since in that case there are additional channels for spin relaxation. That is also the case of impurities that hybridize directly with graphene carbon atoms, such as hydrogen\cite{CastroNetoPaco}. In that case, an enhancement of the SO coupling, $ \Delta_{loc} ( \mathbf{r} )$ is induced due to the local distortion of the lattice coordination. For $\Delta_{loc} \gg \Delta$ it can be shown that $\alpha \propto \langle \Delta_{loc}^2 \rangle / E_{loc}^2$, where $E_{loc}$ is an energy scale comparable to the local shift of the chemical potential in the region where the spin orbit coupling is modified. Alternatively, scattering by local spins can modify significantly the spin relaxation, whose effect could be determined by the dependence on temperature, magnetic field, or injected current. It is worth mentioning that the combination of Zeeman coupling with local moments and enhanced spin-orbit coupling can lead to interesting new effects\cite{Qetal10}.

Another interesting consequence of our work is that, if we suppose weak scatterers so that $\tau_p\propto \epsilon_F^{-1}$, the Elliot-Yafet mechanism implies $\tau_s\propto \epsilon_F$, which scales with carrier density in the same way as $\tau_p^{-1}$. This behavior makes it difficult to distinguish this mechanism and the D'yakonov-Perel one ($\tau_s\propto\tau_p^{-1}$), which could explain the behavior of spin lifetime at high temperatures\cite{HanKawakami}.

{\it Acknowledgments}: We appreciate useful discussions with R. Kawakami, B. J. van Wees, J. Fabian, V. I. Falko, B. Özyilmaz, S. Das Sarma, and A. K. Geim. AHCN acknowledges DOE grant DE-FG02-08ER46512 and ONR grant MURI N00014-09-1-1063. HO acknowledges financial support through grant JAE-Pre (CSIC, Spain). This work was also supported by MICINN (Spain) through grants FIS2008-00124 and
CONSOLIDER CSD2007-00010.

\bibliography{elliotyafet}

\begin{widetext}

\section*{Supplementary information}

\subsection{Spin relaxation within the Born approximation}

Let's consider the scattering problem defined by the Hamiltonian $\hat{\mathcal{H}}_0+U$, where $\hat{\mathcal{H}}_0$ is nothing but the free Hamiltonian of Eq.~(4) in the main text, and $U$ is a generic scattering potential, diagonal in sublattice and spin indices. We take $\Delta=0$ for the moment. We treat the problem in perturbation theory. To first order in the scattering potential (Born approximation), the wave function solution of the problem with energy $\epsilon=\hbar v_Fk$ reads $\Psi^{(0)}+\Psi^{(1)}$, where $\Psi^{(0)}$ is solution in the absence of the scattering potential, and $\Psi^{(1)}$ is given by:\begin{equation}
\Psi^{(1)}\left(\mathbf{r}\right)=\int d^2\mathbf{r}'\hat{G}\left(\epsilon,\mathbf{r}-\mathbf{r}'\right)\left[\hat{\mathcal{H}}_0
+\epsilon\hat{\mathcal{I}}\right]U\left(\mathbf{r}'\right)\Psi^{(0)}\left(\mathbf{r}'\right)
\end{equation}where\begin{equation}
\hat{G}\left(\epsilon,\mathbf{r}\right)=\left[\left(\epsilon+i0\right)^2-\hat{\mathcal{H}}_0^2\right]^{-1}
\label{green}
\end{equation}
In this case, the Green function \eqref{green} has a trivial structure in spin and sublattice indices, $\hat{G}\left(\epsilon,\mathbf{r}\right)=
G\left(\epsilon,\mathbf{r}\right)\hat{\mathcal{I}}$, where it reads:\begin{equation}
G\left(\epsilon,\mathbf{r}\right)=\frac{1}{\left(2\pi\right)^2}\int d^2\mathbf{q}
\frac{e^{i\mathbf{q}\cdot\mathbf{r}}}{\left(\epsilon+i0\right)^2-\left(\hbar v_Fq\right)^2}=-\frac{i}{4\left(\hbar v_F\right)^2}H_0^{(1)}\left(kr\right)
\end{equation}which is nothing but the 2D Klein-Gordon propagator ($H_n^{(1)}$ are the Hankel functions of first kind). The asymptotic form of the Green function is:\begin{equation}
G\left(\epsilon,\mathbf{r}\right)\approx-\frac{\left(\hbar v_F\right)^{-2}}{\sqrt{-i8\pi kr}}\cdot e^{ikr}
\end{equation}In terms of a conventional scattering problem, we take as incoming wave function $\Psi^{in}\equiv\Psi^{(0)}$ a Bloch state with positive helicity. In the light of the asymptotic form of the Green function, it is clear that the scattered wave behaves as $e^{ikr}/\sqrt{-ir}$ in the asymptotic limit. Moreover, as it is sketched in Fig.~1 of the main text, for each scattering angle $\theta$, two different scattered waves must be considered, with opposite helicity. Then, the wave function of the problem can be written in the asymptotic limit as follows:\begin{equation}
\Psi=\Psi^{(0)}+\frac{f_+\left(\theta\right)}{\sqrt{-ir}}\left[\left(\begin{array}{c}
1 \\
e^{i\theta_{\mathbf{k}}}\end{array}\right)\otimes|\uparrow\rangle+i\left(\begin{array}{c}
e^{i\theta_{\mathbf{k}}}\\
e^{2i\theta_{\mathbf{k}}}\end{array}\right)\otimes|\downarrow\rangle\right]e^{ikr}+
\frac{f_-\left(\theta\right)}{\sqrt{-ir}}\left[\left(\begin{array}{c}
1 \\
e^{i\theta_{\mathbf{k}}}\end{array}\right)\otimes|\uparrow\rangle-i\left(\begin{array}{c}
e^{i\theta_{\mathbf{k}}}\\
e^{2i\theta_{\mathbf{k}}}\end{array}\right)\otimes|\downarrow\rangle\right]e^{ikr}
\end{equation}
where $f_h\left(\theta\right)$ is the scattering amplitude at each chiral channel. By assuming the usual approximation $|\mathbf{r}-\mathbf{r}'|\approx r-\mathbf{r}'\cdot\hat{\mathbf{r}}$, and integrating by parts is easy to see that:\begin{equation}\Psi^{(1)}\left(\mathbf{r}\right)=-\frac{\left(\hbar v_F\right)^{-1}}{\sqrt{8\pi k }}\left(\int d^2\mathbf{r}'U\left(\mathbf{r}'\right)e^{i\left(\mathbf{k}
-\mathbf{k}'\right)\cdot\mathbf{r}'}\right)
\left[\vec{\sigma}\cdot\mathbf{k}'+k\mathcal{I}\right]
\cdot\left[\left(\begin{array}{c}
1 \\
e^{i\theta_{\mathbf{k}}}\end{array}\right)\otimes|\uparrow\rangle+i\left(\begin{array}{c}
e^{i\theta_{\mathbf{k}}}\\
e^{2i\theta_{\mathbf{k}}}\end{array}\right)\otimes|\downarrow\rangle\right]\frac{e^{ik r}}{\sqrt{-ir}}
\end{equation}where $\mathbf{k}'=k\hat{\mathbf{r}}$. Then it is straightforward to obtain the scattering amplitudes given in Eq.~(6) of the main text.

In order to compute exactly the scattering amplitudes in the presence of the SO interaction we must add the Rashba-like coupling to $\hat{\mathcal{H}}_0$. The first consequence is that the Green function acquires a non-trivial structure in spin and sublattice degrees of freedom. After some tedious calculus we obtain:\begin{eqnarray}
\hat{G}\left(\epsilon,\mathbf{r}\right)=-\frac{i}{8\left(\hbar v_F\right)^2}\left(H_0^{(1)}\left(k_+r\right)+H_0^{(1)}\left(k_-r\right)\right)\hat{\mathcal{I}}
+\frac{i\Delta}{16\left(\hbar v_F\right)^2\epsilon}\left(H_0^{(1)}\left(k_+r\right)-H_0^{(1)}\left(k_-r\right)\right)\left(\hat{\mathcal{I}}+\hat{\sigma}_z\otimes \hat{s}_z\right)-\nonumber\\
-\frac{i}{8\hbar v_F\epsilon}\left(k_+H_1^{(1)}\left(k_+r\right)-k_-H_1^{(1)}\left(k_-r\right)\right)\left(e^{-i\theta}\hat{\mathcal{I}}
\otimes \hat{s}^+-e^{i\theta}\hat{\mathcal{I}}\otimes \hat{s}^-\right)
\end{eqnarray}where $k_{\pm}=\left(\hbar v_F\right)^{-1}\sqrt{\epsilon^2\mp\epsilon\Delta}$, $\hat{s}^{\pm}=\left(\hat{s}_x\pm i\hat{s}_y\right)/2$, and $\hat{\sigma}_i$ ($\hat{s}_i$) are Pauli matrixes associated to the sublattice (spin) degrees of freedom. We can repeat the same analysis as before. For an incoming wave with energy $\epsilon$ in the positive quasi-helicity channel there are two outgoing radial waves with the same energy, with associated scattering amplitudes:\begin{eqnarray}
f_+^{\Delta}\left(\theta\right)=-\left(\hbar v_F\right)^{-2}\sqrt{\frac{1}{8\pi k_+}}\left(\epsilon+\left(\epsilon-\Delta\right)\cos\theta\right) U_{\mathbf{q}_+}e^{-i\theta}
\nonumber\\
f_-^{\Delta}\left(\theta\right)=-\left(\hbar v_F\right)^{-2}\sqrt{\frac{1}{8\pi k_-}}\left(\epsilon+\Delta\right)U_{\mathbf{q}_-}ie^{-i\theta}\sin\theta
\label{scattering_amplitudes}\end{eqnarray}where $\mathbf{q}_{\pm}=\mathbf{k}'_{\pm}-\mathbf{k}$ is the transferred momentum, with $\mathbf{k}'_{\pm}=k_{\pm}\hat{\mathbf{r}}$. Note that in the presence of the Rashba-like coupling the spin and sublattice degrees of freedoms are completely entangled. As consequence, backscattering is not forbidden, since eigenstates of $\hat{\mathcal{H}}_0$ given by Eq.~(5) of the main text are not longer eigenstates of the sublattice chirality operator $\frac{\vec{\sigma}\cdot\mathbf{k}}{2k}$. In fact:\begin{equation}
\left[\left(\vec{\sigma}\times\vec{s}\right)_z,\frac{\vec{\sigma}\cdot\mathbf{k}}{2k}\right]=
i\frac{\vec{s}\cdot\mathbf{k}}{k}\sigma_z
\end{equation}

The results for weak and Coulomb scatterers were obtained by evaluating numerically Eqs.~(7)-(8) of the main text with the expressions deduced here.

\subsection{Spin relaxation for scattering centers with cylindrical symmetry}

\subsubsection{Strong scatterers}

As it is well known, scattering by strong scatterers such as vacancies or resonant impurities cannot be studied within the Born approximation. Strong scatterers are used to be described as infinite potentials with a range of the order of the lattice constant. We assume isotropy in order to exploit the cylindrical symmetry of the problem by using the decomposition of the eigenstates into partial waves with well-defined generalized total angular momentum $J=l_z+\sigma_z/2+s_z/2$, which is actually a global symmetry of the problem, where $l_z$ is third component of the orbital angular momentum operator $l_z=-i\left(x\partial_y-y\partial_x\right)$. For each incoming wave with energy $\epsilon$, there are two reflected waves with the same energy, characterized by the reflection coefficients $r_1^l$, $r_2^l$, where $l$ labels the angular momentum channel $J=l+1$. Then, following \citep{HuertasHernandoetalprl}, we define:\begin{equation}
S=\frac{\sum_{l}\left(\left|r_1^{0l}\right|\cdot\left|r_1^{l}-r_1^{0l}\right|+
\left|r_2^{0l}\right|\cdot\left|r_2^{l}-r_2^{0l}\right|\right)}
{\sum_{l}\left(\left|r_1^{0l}\right|^2+\left|r_2^{0l}\right|^2\right)}
\label{s_num}
\end{equation}where $r_{1,2}^{0l}$ is nothing but $r_{1,2}^{l}$ in the absence of SO coupling. This quantity can be seen as the amount of spin relaxation during a scattering event.

Neglecting mixing of the two inequivalent valleys, an incoming wave with energy $\epsilon$ and total angular momentum $J=l+1$ can be written as:
\begin{equation}
\Psi_{in}=\left(\begin{array}{c}
J_{l}\left(k_+r\right)e^{il\theta}\\
ic_+J_{l+1}\left(k_+r\right)e^{i\left(l+1\right)\theta}\end{array}\right)\otimes|\uparrow\rangle-
\left(\begin{array}{c}
c_+J_{l+1}\left(k_+r\right)e^{i\left(l+1\right)\theta}\\
iJ_{l+2}\left(k_+r\right)e^{i\left(l+2\right)\theta}\end{array}\right)\otimes|\downarrow\rangle
\label{incoming}
\end{equation}
where $c_{\pm}=\frac{\epsilon}{\hbar v_Fk_{\pm}}$ and $k_{\pm}$ is defined as above. The outgoing wave can be written as the superposition:
\begin{eqnarray}
\Psi_{out}=r_{1}^l\left[\left(\begin{array}{c}
Y_{l}\left(k_+r\right)e^{il\theta}\\
ic_+Y_{l+1}\left(k_+r\right)e^{i\left(l+1\right)\theta}\end{array}\right)\otimes|\uparrow\rangle-
\left(\begin{array}{c}
c_+Y_{l+1}\left(k_+r\right)e^{i\left(l+1\right)\theta}\\
iY_{l+2}\left(k_+r\right)e^{i\left(l+2\right)\theta}\end{array}\right)\otimes|\downarrow\rangle\right]+\nonumber\\
+r_{2}^l\left[\left(\begin{array}{c}
Y_{l}\left(k_-r\right)e^{il\theta}\\
ic_-Y_{l+1}\left(k_-r\right)e^{i\left(l+1\right)\theta}\end{array}\right)\otimes|\uparrow\rangle+
\left(\begin{array}{c}
c_-Y_{l+1}\left(k_-r\right)e^{i\left(l+1\right)\theta}\\
iY_{l+2}\left(k_-r\right)e^{i\left(l+2\right)\theta}\end{array}\right)\otimes|\downarrow\rangle\right]
\label{outcoming}
\end{eqnarray}

We model a strong scatterer as a circular void of radius $R$. By imposing zig-zag boundary conditions at r=R, we obtain the following expressions for the reflection coefficients at each $l$ channel:\begin{eqnarray}
r_1^l=-\frac{k_-J_{l+1}\left(k_+R\right)Y_{l}\left(k_-R\right)+k_+J_{l}\left(k_+R\right)Y_{l+1}\left(k_-R\right)}
{k_+Y_{l}\left(k_+R\right)Y_{l+1}\left(k_-R\right)+k_-Y_{l}\left(k_-R\right)Y_{l+1}\left(k_+R\right)}\label{formula1}\\
r_2^l=\frac{k_-J_{l+1}\left(k_+R\right)Y_{l}\left(k_+R\right)-k_-J_{l}\left(k_+R\right)Y_{l+1}\left(k_+R\right)}
{k_+Y_{l}\left(k_+R\right)Y_{l+1}\left(k_-R\right)+k_-Y_{l}\left(k_-R\right)Y_{l+1}\left(k_+R\right)}\label{formula2}
\end{eqnarray}At low energies $kR\ll1$, $S$ is dominated by the first harmonics $l=-1,0$. In fact, the following asymptotic expression is deduced:\begin{equation}
S\approx\frac{\pi}{2}\cdot\frac{\Delta}{\epsilon_F}
\label{s_analitic}
\end{equation}which fits very well the exact result, as it is shown in Fig.~3 of the main text.

\subsubsection{Clusters of impurities}

In the case of weak scatterers, if the range of the scattering potential $R$ is too large in such a way that the associated energy scale $\hbar v_F R^{-1}$ exceeds its strength, $VR\ll\hbar v_F$, then the Born approximation fails. We employ the same model as in the case of strong scatterers. The incoming and outgoing waves are \eqref{incoming} and \eqref{outcoming} respectively. Inside the potential $r<R$, the wave function regular at the origin can be written as the superposition:
\begin{eqnarray}
\Psi_{inside}=t_{1}^l\left[\left(\begin{array}{c}
J_{l}\left(q_+r\right)e^{il\theta}\\
ic_+'J_{l+1}\left(q_+r\right)e^{i\left(l+1\right)\theta}\end{array}\right)\otimes|\uparrow\rangle-
\left(\begin{array}{c}
c_+'J_{l+1}\left(q_+r\right)e^{i\left(l+1\right)\theta}\\
iJ_{l+2}\left(q_+r\right)e^{i\left(l+2\right)\theta}\end{array}\right)\otimes|\downarrow\rangle\right]+\nonumber\\
+t_{2}^l\left[\left(\begin{array}{c}
J_{l}\left(q_-r\right)e^{il\theta}\\
ic_-'J_{l+1}\left(q_-r\right)e^{i\left(l+1\right)\theta}\end{array}\right)\otimes|\uparrow\rangle+
\left(\begin{array}{c}
c_-'J_{l+1}\left(q_-r\right)e^{i\left(l+1\right)\theta}\\
iJ_{l+2}\left(q_-r\right)e^{i\left(l+2\right)\theta}\end{array}\right)\otimes|\downarrow\rangle\right]
\label{inside}
\end{eqnarray}where $q_{\pm}=\sqrt{\left(\epsilon-V\right)^2\mp\Delta \left(\epsilon-V\right)}/\left(\hbar v_F\right)$ and $c_{\pm}'=\frac{\epsilon-V}{\hbar v_Fq_{\pm}}$

\begin{figure}
\begin{center}
\includegraphics[width=0.5\textwidth]{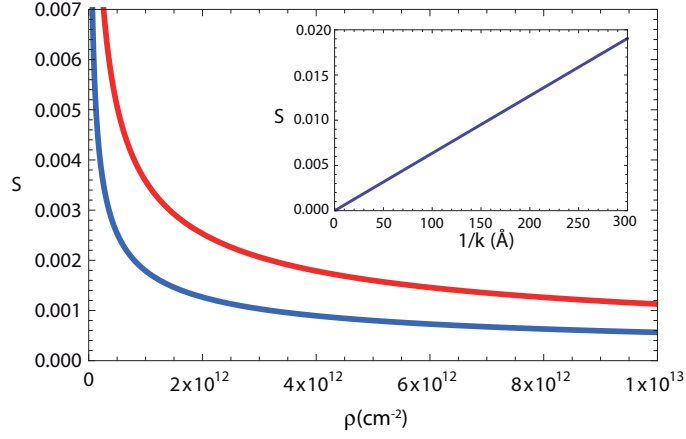}
\end{center}
\caption{Weak scatterers: $S$ as a function of the carrier concentration for $\Delta=1$ meV (in blue) and $\Delta=0.5$ meV (in red), in both cases $R=1$ {\AA} and $V_0=0.1$ eV. These results are obtained by solving numerically the system \eqref{system}. Inset: Dependence on $k_F^{-1}$, $\Delta=0.5$ meV.}
\label{weak}
\end{figure}

By imposing matching conditions at $r=R$ we obtain 4 equations that define the unknowns of the problem, in particular $r_1^l$ and $r_2^l$:\begin{eqnarray}
J_{l}\left(k_+R\right)+r_1^lY_{l}\left(k_+R\right)+r_2^lY_{l}\left(k_-R\right)=
t_1^lJ_{l}\left(q_+R\right)+t_2^lJ_{l}\left(q_-R\right)\nonumber\\
c_+J_{l+1}\left(k_+R\right)+r_1^lc_+Y_{l+1}\left(k_+R\right)+r_2^lc_-Y_{l+1}\left(k_-R\right)=
t_1^lc_+'J_{l+1}\left(q_+R\right)+t_2^lc_-'J_{l+1}\left(q_-R\right)\nonumber\\
c_+J_{l+1}\left(k_+R\right)+r_1^lc_+Y_{l+1}\left(k_+R\right)-r_2^lc_-Y_{l+1}\left(k_-R\right)=
t_1^lc_+'J_{l+1}\left(q_+R\right)-t_2^lc_-'J_{l+1}\left(q_-R\right)\nonumber\\
J_{l+2}\left(k_+R\right)+r_1^lY_{l+2}\left(k_+R\right)-r_2^lY_{l+2}\left(k_-R\right)=
t_1^lJ_{l+2}\left(q_+R\right)-t_2^lJ_{l+2}\left(q_-R\right)
\label{system}\end{eqnarray}In the absence of SO coupling, the reflection coefficients read ($\xi\equiv sgn\left(\epsilon-V\right)$):\begin{eqnarray}
r_1^{0l}=\frac{1}{2}\left[\frac{\xi J_l\left(qR\right)J_{l+1}\left(kR\right)-J_l\left(kR\right)J_{l+1}\left(qR\right)}{Y_{l}\left(kR\right)J_{l+1}\left(qR\right)-\xi Y_{l+1}\left(kR\right)J_{l}\left(qR\right)}+\frac{\xi J_{l+1}\left(qR\right)J_{l+2}\left(kR\right)-J_{l+1}\left(kR\right)J_{l+2}\left(qR\right)}{Y_{l+1}\left(kR\right)J_{l+2}\left(qR\right)-\xi Y_{l+2}\left(kR\right)J_{l+1}\left(qR\right)}\right]\nonumber\\
r_2^{0l}=\frac{1}{2}\left[\frac{\xi J_l\left(qR\right)J_{l+1}\left(kR\right)-J_l\left(kR\right)J_{l+1}\left(qR\right)}{Y_{l}\left(kR\right)J_{l+1}\left(qR\right)-\xi Y_{l+1}\left(kR\right)J_{l}\left(qR\right)}-\frac{\xi J_{l+1}\left(qR\right)J_{l+2}\left(kR\right)-J_{l+1}\left(kR\right)J_{l+2}\left(qR\right)}{Y_{l+1}\left(kR\right)J_{l+2}\left(qR\right)-\xi Y_{l+2}\left(kR\right)J_{l+1}\left(qR\right)}\right]
\end{eqnarray}

At low energies $k_FR\ll1$, the most relevant channels correspond to $l=0,-1,-2$. These reflection coefficients behave as $r\sim\frac{V\epsilon R^2}{\left(\hbar v_F\right)^2}$. In order to estimate the effect of the SO coupling in the doped regime it is enough to consider the energy splitting of the chiral sub-bands, as it is argued in the main text. The substitution $\epsilon\rightarrow\epsilon\pm\Delta/2+O\left(\Delta\right)^2$ gives $r_1\approx \frac{V\epsilon R^2}{\left(\hbar v_F\right)^2}+\frac{\Delta VR^2}{2\left(\hbar v_F\right)^2}+O\left(\Delta/\epsilon\right)^2$ and $r_2\approx \frac{V\epsilon R^2}{\left(\hbar v_F\right)^2}-\frac{\Delta VR^2}{2\left(\hbar v_F\right)^2}+O\left(\Delta/\epsilon\right)^2$. So the averaged amount of relaxed spin behaves as $S\sim\Delta/\epsilon$, as it is numerically shown in Figure \ref{weak}. This is the same result ($S$ is in fact of the same order) as the one deduced within the Born approximation.

The opposite limit $k_FR\gg1$ has been studied in order to analyze scattering by clusters of impurities \cite{KatsnelsonPacoGeim}. Assuming that charged impurities inside the cluster do not break the sublattice symmetry, so there is no gap opening, the main effect is a local shift of the chemical potential $\epsilon_F\rightarrow\epsilon_F+V$ ($V>0$, note the change in the sign of $V$ in relation with the previous calculation) inside the cluster. The effect of the long-range Coulomb potential has been analyzed before. The present approach is valid for circular shapes, where $R$ is the radius of the cluster. For clusters large enough we can consider the regime characterized by $kR\gg1$. If we take into account the asymptotic behavior of the Bessel functions:\begin{eqnarray}
J_l\left(x\right)\approx\sqrt{\frac{2}{\pi x}}\cos\left(x-\frac{n\pi}{2}-\frac{\pi}{4}\right)\nonumber\\
Y_l\left(x\right)\approx\sqrt{\frac{2}{\pi x}}\sin\left(x-\frac{n\pi}{2}-\frac{\pi}{4}\right)
\end{eqnarray}it is easy to see that in this regime $r_2\sim 0$ and:\begin{equation}
r_1\approx\tan\left(Rk-Rq\right)=-\tan\left(\frac{VR}{\hbar v_F}\right)
\end{equation}
\begin{figure}
\begin{center}
\includegraphics[width=0.5\textwidth]{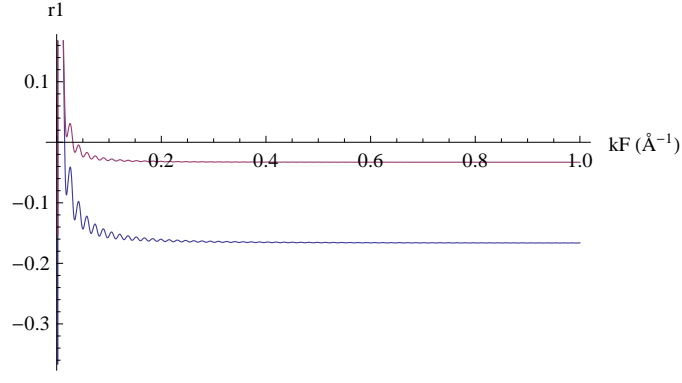}
\end{center}
\caption{The reflection coefficient $r_1$ in the limit $k_FR\gg1$ for $V=0.1$ eV and $V=0.5$ eV, and $R=20$ nm. $r_1$ tends to the asymptotic value $-\tan\left(\frac{VR}{\hbar v_F}\right)$, -0.033 and -0.167 respectively, as $k_F$ grows.}
\label{oscillations}
\end{figure}
As it is showed in Fig. \ref{oscillations}, in this regime  $r_1$  as a function of $k$ presents some oscillations which attenuate as $k$ grows, approaching the asymptotic value $-\tan\left(VR/\hbar v_F\right)$. These oscillations translate to the scattering cross section \cite{KatsnelsonPacoGeim}, and, importantly, to $S$. However, these oscillations are expected to be averaged out for less symmetric cluster shapes, as it is deduced from semi-classical arguments \cite{KatsnelsonPacoGeim}. In order to do so within our approach, we study the system $\eqref{system}$ in the limit $k_{\pm}R\gg1$. By taking the asymptotic form of the Bessel functions we arrive to:\begin{equation}
r_1^{l}\approx\frac{c_+\cos\left(q_+R-\frac{\pi}{4}-
\frac{l\pi}{2}\right)\sin\left(k_+R-\frac{\pi}{4}-\frac{l\pi}{2}\right)-c_+'\cos\left(k_+R-\frac{\pi}{4}-
\frac{l\pi}{2}\right)\sin\left(q_+R-\frac{\pi}{4}-\frac{l\pi}{2}\right)}{c_+\cos\left(q_+R-\frac{\pi}{4}-
\frac{l\pi}{2}\right)\cos\left(k_+R-\frac{\pi}{4}-\frac{l\pi}{2}\right)+c_+'\sin\left(q_+R-\frac{\pi}{4}-
\frac{l\pi}{2}\right)\sin\left(k_+R-\frac{\pi}{4}-\frac{l\pi}{2}\right)}
\end{equation}Then, we define the amount of spin relaxation as:\begin{equation}
S=\frac{\sum_l\left|r_1^{0l}\right|\cdot\left|r_1^l-r_1^{0l}\right|}{\sum_l\left|r_1^{0l}\right|^2}
\end{equation}
The results are shown in Figs. \ref{eyc1} and \ref{eyc2}. The oscillations of $S$ as a function of $k_F$ persist, but clearly the enveloping curve behaves as $\sim1/k_F$. So $S\sim\Delta/\epsilon_F$, as the previous cases.

\begin{figure}
\begin{center}
\includegraphics[width=0.5\textwidth]{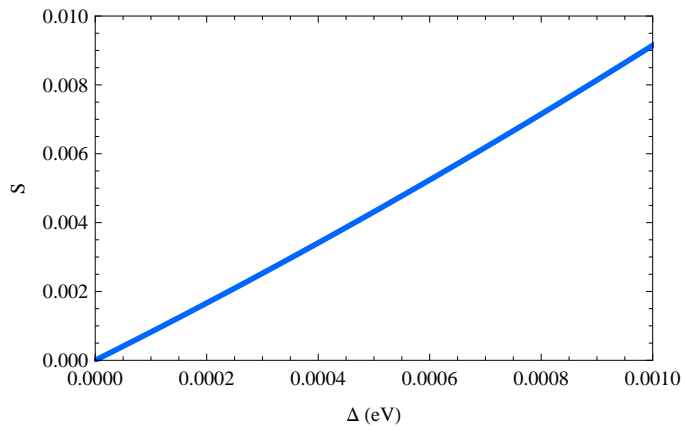}
\end{center}
\caption{$S$ as a function of $\Delta$ in the case of scattering by clusters of impurities for $V=0.5$ eV, $R=20$ nm and $k_F=0.01$ {\AA}$^{-1}$.}
\label{eyc1}
\end{figure}

\begin{figure}
\begin{center}
\includegraphics[width=0.5\textwidth]{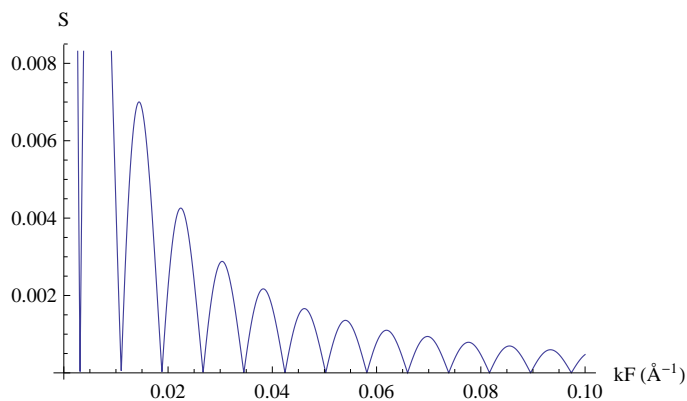}
\end{center}
\caption{$S$ as a function of $k_F$ in the case of scattering by clusters of impurities for $V=0.5$ eV, $R=20$ nm and $\Delta=0.5$ meV.}
\label{eyc2}
\end{figure}

\subsection{Effect of a local enhancement of the SO coupling}

As it is argued in the main text, if the scatterer induces a local enhancement of the SO coupling, then the Elliot relation does not hold since there are additional channels for spin relaxation apart from the one induced by momentum scattering. Since the source is the same, the presence of the scatterer, a correlation between the momentum relaxation time and the spin relaxation time is expected. A quantity $\alpha$ can be defined, like in the main text, as the spin-flip probability during a momentum scattering event. Nevertheless, two different mechanisms contribute to $\alpha$ in this case, the spin-flip induced by momentum relaxation (Elliot-Yafet), and the spin-flip induced by the local SO coupling. The latter manifests itself in the calculation within the Born approximation as a new contribution to the scattering amplitude in the positive helicity channel. If we assume a local enhancement of the SO coupling near the scatterer $\Delta^{loc}\left(\mathbf{r}\right)\left(\vec{\sigma}\times \vec{s}\right)_z$ as a perturbation to $\hat{\mathcal{H}}_0$, then we obtain this new contribution to scattering in the positive helicity channel given by:\begin{equation}
f_+^{\Delta^{loc}}\left(\theta\right)=-\left(\hbar v_F\right)^{-2}\sqrt{\frac{1}{8\pi k_+}}\epsilon \Delta^{loc}_{\mathbf{q}_+}e^{-i\theta}
\end{equation}where $\Delta^{loc}_{\mathbf{q}_+}$ is the Fourier transformation of the local coupling evaluated at the transferred momentum $\mathbf{q}_+$ defined as before.

If we assume that $\Delta^{loc}\gg\Delta$, as it is argued in the main text, then we can neglect the Elliot-Yafet and study separately this new mechanism. We take $\Delta=0$ for convenience, since then the spin up and spin down channels are asymptotically well defined. If we consider an incoming Bloch state in the spin up channel with energy $\epsilon=\hbar v_F k$, then the scattering amplitudes in the spin up and spin down channels read:\begin{eqnarray}
f_{\uparrow}\left(\theta\right)=-\left(\hbar v_F\right)^{-1}\sqrt{\frac{k}{8\pi}}U_{\mathbf{q}}\left(1+e^{-i\theta}\right)
\nonumber\\
f_{\downarrow}\left(\theta\right)=-\left(\hbar v_F\right)^{-1}\sqrt{\frac{k}{8\pi}}\Delta^{loc}_{\mathbf{q}}ie^{-i\theta}
\label{scattering_amplitudes}\end{eqnarray}
Clearly $\alpha\propto\left|\frac{f_{\downarrow}}{f_{\uparrow}}\right|^2$. If one assumes that the scattering center consists on a finite region where both a local shift of the chemical potential of the order of $E_{loc}$ and an enhancement of the SO coupling of the order of $\Delta_{loc}$ are induced, then $\alpha\propto\Delta_{loc}^2/E_{loc}^2$, as it is pointed out in the main text. That is the case of heavy impurities. The analysis is more complicated in the case of resonant impurities which induces a local enhancement of the SO coupling due to the local distortion of the lattice coordination. The Born approximation fails in that case. By considering a model as the one described in \citep{AntonioPaco} and including also a local shift of the chemical potential, then it can be shown that the scattering cross section has a non-monotonic behavior as a function of the carrier concentration. In other words, different regimes must be taken into account and the previous statement is not so solid. Another interesting point is that in the case of resonant impurities the spatial decay of the local SO coupling is governed by the dispersion relation of $\sigma$ bands \citep{AntonioPaco}. To consider a step function for this coupling is probably a crude approximation which influences on the dependence of the spin relaxation time on the carrier concentration.

\end{widetext}

\end{document}